\documentclass[]{scrartcl}

\let\proglang=\textsf
\newcommand{\pkg}[1]{{\fontseries{m}\fontseries{b}\selectfont #1}}

\usepackage{geometry}
\usepackage{amsmath, amsfonts, amssymb, bm}
\usepackage{amsthm}
\usepackage{thumbpdf,lmodern}  
\usepackage{placeins}

\usepackage[dvipsnames]{xcolor}
\definecolor{darkteal}{RGB}{0,102,102}
\definecolor{darkerteal}{RGB}{0,76,76}

\usepackage[colorlinks = true,
linkcolor = Mahogany,
urlcolor  = Mahogany,
citecolor = Mahogany,
anchorcolor = Mahogany]{hyperref}

\usepackage{natbib}  
\newcommand*{\doi}[1]{\href{http://dx.doi.org/#1}{\textsc{doi:} #1}}

\DeclareOldFontCommand{\bf}{\normalfont\bfseries}{\mathbf}
\usepackage{comment}  
\usepackage{bbm,dsfont}  
\usepackage{verbatim}  
\usepackage{tabularx}  
\newcolumntype{s}{>{\hsize=.30\hsize}X}
\newcolumntype{z}{>{\hsize=.15\hsize}X}
\usepackage{booktabs}  
\usepackage[english]{babel}
\usepackage{graphicx}  
\usepackage{pifont}  

\title{\texttt{sanba}: An R Package for Bayesian Clustering of Distributions via Shared Atoms Nested Models}
\author{Francesco Denti\footnote{University of Padova, \texttt{francesco.denti@unipd.it}} $\:$ and Laura D'Angelo\footnote{University of Milan-Bicocca, \texttt{laura.dangelo@unimib.it}}}
\date{}

\begin{document}

\maketitle

\begin{abstract}
Nested data structures arise when observations are grouped into distinct units, such as patients within hospitals or students within schools. Accounting for this hierarchical organization is essential for valid inference, as ignoring it can lead to biased estimates and poor generalization. This article addresses the challenge of clustering both individual observations and their corresponding groups while flexibly estimating group-specific densities. Bayesian nested mixture models offer a principled and robust framework for this task. However, their practical use has often been limited by computational complexity. To overcome this barrier, we present \pkg{sanba}, an \proglang{R} package for Bayesian analysis of grouped data using nested mixture models with a shared set of atoms, a structure recently introduced in the statistical literature. The package provides multiple inference strategies, including state-of-the-art Markov Chain Monte Carlo routines and variational inference algorithms tailored for large-scale datasets. All core functions are implemented in \proglang{C++} and seamlessly integrated into \proglang{R}, making \pkg{sanba} a fast and user-friendly tool for fitting nested mixture models with modern Bayesian algorithms.

\end{abstract}

\section{Introduction}
\label{sec::introduction}

Nested data, where individual observations are organized into separate,
related groups, arise frequently in experimental and observational
studies and are commonly analyzed using hierarchical (nested) models.
These models represent a good compromise between the cases of complete
pooling, obtained when combining all observations into a unique sample,
and complete independence, obtained when modeling each group separately.
By coherently integrating information across groups, hierarchical models
enable the integration of information across groups while preserving
group-level heterogeneity. However, standard hierarchical parametric
models, where a parametric density is assumed for the group-specific
distributions, generally limit the borrowing of information to the
specific aspects of the distribution captured by the parameters. This
can lead to suboptimal performance when the underlying distributions
exhibit complex features such as multimodality, skewness, or heavy
tails. In such settings, simple parametric models may fail to adequately
represent the data. A more flexible approach is provided by Bayesian
parametric and nonparametric mixtures. These models have shown strong
performance in modeling data with non-standard distributions. In recent
years, a growing body of literature has extended mixture models beyond
the classical exchangeable setting, leading to the development of more
flexible tools for complex data structures. While our focus is on nested
designs, related advances have addressed scenarios involving covariates,
censoring, multiple data sources, or spatial and temporal dependencies
among variables \citep[see, e.g.,][]{lit01, lit02, lit03, Chandra2023}.

Mixture models are widely used not only for flexible density estimation
but also for uncovering latent structure through clustering. In the
context of nested data, the clustering objective can be twofold: one may
wish to identify subgroups of individuals with similar characteristics
or detect group-level clusters sharing similar outcome distributions.
This type of analysis has been extensively studied within the Bayesian
nonparametric (BNP) framework, particularly through models based on
hierarchical constructions of Dirichlet processes (DP). Notable examples
include the nested DP \citep{Rodriguez2008}, the semi-hierarchical DP
\citep{beraha2021}, the common atoms model \citep[CAM;][]{Denti2023},
its finite counterpart \citep{DAngelo2023} and generalization
\citep{dentidangelo2025}, as well as the hidden hierarchical DP
\citep{lijoi2023} and the shared atoms nested models \citep[SAN;][]{SAN}.

The interest in Bayesian mixture models is evident not only in
theoretical research but also among practitioners, as reflected in the
increase of packages and software designed to simplify their application
for non-specialists. In the context of Bayesian finite mixture models,
examples include the \pkg{bayesmix} package \citep{bayesmix}, which fits
finite mixtures of univariate Gaussian distributions using \pkg{JAGS};
the {\proglang{R}}{} package \pkg{bmixture} \citep{bmixture}, which
supports Gaussian, Student's t, and gamma mixtures; and the
{\proglang{R}}{} package \pkg{mixAK} \citep{KOMAREK}, which provides
model-based clustering for longitudinal data. For nonparametric
mixtures, notable {\proglang{R}}{} packages are \pkg{bspmma}
\citep{bspmma}, which implements a semi-parametric random effects model
for meta-analysis; \pkg{PReMiuM} \citep{premium_pkg}, for nonparametric
regression using covariate-dependent DP mixtures; \pkg{BNPmix}
\citep{bnpmix_pkg}, which handles nonparametric density estimation and
regression via Pitman--Yor mixtures; and \pkg{dirichletprocess}
\citep{Ross2019dirichletprocess}, which offers flexible tools to
facilitate MCMC inference for user-specified DP mixture models. Beyond
the {\proglang{R}}{} environment, the \proglang{C++} library
\pkg{BayesMix} \citep{beraha2022bayesmix} implements various algorithms
for MCMC posterior simulation in flexible BNP mixtures. Finally, it is
worth noting the software \pkg{JAGS} \citep{JAGS} and the
{\proglang{R}}{} package \pkg{NIMBLE} \citep{nimble}, whose scope is
broader than the previous examples, and which also support the inclusion
of certain nonparametric mixture priors.

All the packages mentioned above rely on MCMC algorithms to perform
inference. These algorithms are well-established and deliver reliable
results, even for highly complex models. However, simulation-based
methods do not scale well as data dimensionality increases. This
limitation often discourages the use of Bayesian techniques in
large-scale applications. As a result, recent research has increasingly
focused on developing approximate inference algorithms to enable fast
and scalable posterior computation in complex Bayesian models. Among
these, variational inference (VI) stands out as a particularly
successful strategy. Simple Gaussian mixture models estimated via VI are
available, for example, through the \pkg{BayesianGaussianMixture} class
in \pkg{Scikit-learn} \citep{scikitlearn}. The theoretical foundations
of VI methods for BNP mixture models have expanded rapidly, thanks to
seminal contributions such as \citet{Blei2006}, \citet{VB01}, and
\citet{VB02}. However, the number of libraries implementing VI for BNP
models beyond basic Dirichlet process mixtures remains limited. Notable
exceptions include the \proglang{Python} package \pkg{bnpy}
\citep{bnpy}, which supports both MCMC and VI for mixture models, hidden
Markov models, and admixture models---including the hierarchical
Dirichlet process of \citet{Teh2006}; and the {\proglang{R}}{} package
\pkg{mixdir} \citep{Ahlmann-Eltze2019}, which implements hierarchical DP
mixtures for high-dimensional categorical data.

At the time of writing, the available software for implementing BNP
models for nested data remains limited, often confined to GitHub
repositories. The \pkg{{sanba}}{} {\proglang{R}}{} package \citep{sanba}
goes in the direction of filling the gap between recent methodological
advances and the availability of practical, user-friendly tools for
applied users. Specifically, \pkg{sanba} allows for the seamless
application of both MCMC and VI algorithms for posterior inference under
the SAN priors introduced by \citet{SAN} and the CAM model of
\citet{Denti2023}, using mixtures of univariate Gaussian kernels. These
models are particularly suited for application purposes, as they provide
a flexible modeling framework without overlooking the need for a clear
interpretation of the results and scalable computational strategies. To
enhance computational efficiency, all algorithms are implemented in
\proglang{C++} and integrated into {\proglang{R}}{} using \pkg{Rcpp}
\citep{rcpp} and \pkg{RcppArmadillo} \citep{rcpparma}.

The \pkg{sanba} package consolidates and streamlines the content of two
earlier packages in this line of work: \pkg{SANple} \citep{SANplePack}
and \pkg{SANvi} \citep{SANviPack}, both openly available on the
Comprehensive R Archive Network
(CRAN)\footnote{see \url{https://CRAN.R-project.org/package=SANple} and \url{https://CRAN.R-project.org/package=SANvi}}.
\pkg{sanba} attempts to unify the approaches in a unique, user-friendly,
and intuitive suite. The package (at the time of writing, in its version
0.0.2) can be installed from CRAN by

	\begin{verbatim}
		R> install.package("sanba")
	\end{verbatim}

Future stable versions of the package will be available on CRAN, while
development versions can be accessed from the corresponding GitHub
repository and installed using:

	\begin{verbatim}
		R> remotes::install_github("fradenti/sanba")
	\end{verbatim}

The remainder of the paper is organized as follows.
Section\nobreakspace{}\ref{sec::model_and_comput} presents the
methodological background on nested mixture models based on common and
shared atoms. Section\nobreakspace{}\ref{sec::package} illustrates how
to use the package to perform a full analysis of nested data employing a
simple simulated dataset. The main routines of the package are
illustrated, and their output is thoroughly commented on.
Section\nobreakspace{}\ref{sec:beer} presents a real-data analysis,
where we apply a SAN model to a large dataset containing hundreds of
thousands of beer reviews using a variational inference approach.
Section\nobreakspace{}\ref{sec:conclusions} discusses some concluding
remarks and potential future developments.

\section{Nested mixture priors for grouped data}
\label{sec::model_and_comput}
\subsection{Model specification}
\label{sec::model}

The standard framework for Bayesian density estimation is based on the
assumption that the data are exchangeable. However, this assumption can
be restrictive when the outcome is collected under $J$ distinct but
related conditions. In this case, a more suitable assumption is that the
data are \textit{partially exchangeable} (i.e., they are exchangeable
within the groups and conditionally independent across groups). In other
words, observations are supposed to be generated by a collection of
$J$ dependent distributions. Formally, let us denote the vector of
data as $\boldsymbol{y}= (\boldsymbol{y}_1,\dots,\boldsymbol{y}_J)$,
with $\boldsymbol{y}_j$ arising from condition $j$, for
$j=1,\ldots,J$. Within each group, $N_j$ observations are measured:
$\boldsymbol{y}_j=(y_{1,j},\dots,y_{N_j,j})$,
$y_{i,j}\in\mathcal{Y}$, with $\mathcal{Y}$ their support. We adopt
the following hierarchical mixture model specification for
$j=1,\dots,J$: \begin{equation}
	y_{1,j},\dots,y_{N_j,j} \mid f_j \overset{ind.}{\sim} f_j, \quad \quad
	f_j(\cdot) = \int_{\Theta} p(\cdot\mid \theta) G_j(d\theta),
	\label{eq::intro:mixture_model}
\end{equation} where $p(\cdot\mid\theta)$ denotes a parametric kernel
on $\mathcal{Y}\times \Theta$, which depends on the specific nature of
the data, and $\Theta$ is the space of the mixing parameter
$\theta$. Each $G_j$ is the group-specific mixing measure, assumed
to be random and drawn from an almost surely discrete distribution
$Q$, which, therefore, is a \emph{distribution over distributions}.
Formally, we write: \begin{equation*}
	G_1,\dots,G_j \mid Q \sim Q, \quad \quad
	Q = \sum_{k= 1}^{K} \pi_k \delta_{G^*_k}
	\label{eq::intro:nested_prior}
\end{equation*} with $K\in \mathbb{N}\cup \{+\infty\}$. Then, each
\textit{distributional atom} $G^*_k$ has the following form
\begin{equation*}
	G^*_k = \sum_{l= 1}^{L}\omega_{l,k}\delta_{\theta^*_l}, \label{eq::intro:common_atoms}
\end{equation*} with $L\in \mathbb{N}\cup \{+\infty\}$. Finally, the
\textit{observational atoms} $\{\theta^*_l\}_{l= 1}^L$ are randomly
sampled from a non-atomic base measure $H$ on $\Theta$. This
hierarchical construction induces two levels of clustering: a clustering
of groups (referred to as \textit{distributional clustering}) and a
clustering of observations (termed \textit{observational clustering}).
First, we notice that the discreteness of the distribution $Q$ implies
that the outcome in two groups can be modeled using the same
distributional atom $G^*_k$, hence leading to a clustering of the
groups. Then, each distributional atom is itself discrete, allowing two
observations to be assigned to the same atom $\theta^*_l$. Notably,
due to the shared nature of the atoms $\{\theta^*_l\}_{l= 1}^L$ across
the distributions, the clustering of observations is not limited to
being distributional-cluster specific. That is, the same parameter value
can be shared by observations belonging to different groups, regardless
of their distributional cluster assignment, leading to possible
cross-group observational clustering. This nested structure was
originally proposed by \citet{Denti2023} with the CAM. All the models
implemented in the \pkg{sanba} package are based on this formulation and
differ in the distributions of the distributional and observational
weights, respectively, $\{\pi_k\}_{k=1}^{K}$ and
$\{\omega_{l,k}\}_{l=1}^{L}$ for $k=1,\dots,K$.

Specifically, the CAM assumes that both $K$ and $L$ are infinite and
assigns Dirichlet process priors to both the distributional and
observational weights, \begin{equation*}
	\begin{aligned}
		\boldsymbol{\pi}= \{\pi_k\}_{k\geq 1} &\sim \mathrm{GEM}(\alpha)\\
		\boldsymbol{\omega}_k = \{\omega_{l,k}\}_{l\geq 1} &\sim \mathrm{GEM}(\beta), \quad  k \geq 1,    
	\end{aligned}
\end{equation*} where $\mathrm{GEM}(c)$ indicates the Griffiths,
Engen, and McCloskey distribution with parameter $c>0$, which
describes the distribution of DP weights. Additionally, gamma prior
distributions are assumed for the concentration parameters,
$\alpha \sim\mathrm{Gamma}(h^{\alpha}_1,h^{\alpha}_2)$ and
$\beta \sim\mathrm{Gamma}(h^{\beta}_1,h^{\beta}_2)$.

The finite-infinite SAN model \citep[FISAN, ][]{SAN} adopts a ``hybrid''
formulation based on a nonparametric prior for $\boldsymbol{\pi}$ at
the distributional level, and a finite symmetric Dirichlet of fixed
dimension $L$ at the observational level: \begin{equation*}
	\boldsymbol{\omega}_k = (\omega_{1,k},\ldots,\omega_{L,k}) \sim \mathrm{Dirichlet}_L(b, \ldots, b), \quad  k =1,\ldots, K.
\end{equation*} This key modification allows spreading the mass
uniformly over the $L$ components, improving the accuracy of the
estimates for groups with a limited sample size. Following \citet{SAN},
the term \emph{shared} atoms is used instead of \emph{common} atoms to
highlight the distinction with the CAM, where the atoms are implicitly
ordered by the GEM distribution.

Finally, the finite SAN prior
\citep[FSAN, originally proposed in][]{DAngelo2023} assumes that both
sequences of probabilities are finite and they are assigned symmetric
Dirichlet distributions; hence, for the distributional weights, it
assumes \begin{equation*}
	\boldsymbol{\pi}= (\pi_{1},\ldots,\pi_K) \sim \mathrm{Dirichlet}_K(a, \ldots, a).
\end{equation*}

Table\nobreakspace{}\ref{tab::functions} summarizes the implemented
models and the corresponding prior distributions on the two sets of
weights. The choice between parametric or nonparametric priors for the
mixture weights can affect posterior inference since they imply
different properties of the resulting model. A thorough discussion can
be found in \citet{Denti2023}, \citet{SAN}, and
\citet{dentidangelo2025}.

To complete the model formulation, the mixture kernel
in\nobreakspace{}\eqref{eq::intro:mixture_model} is assumed to be
univariate normal, i.e.,
$p(\cdot\mid\theta) = \phi(\cdot\mid\mu,\sigma^2)$, where
$\theta = (\mu,\sigma^2)$ and
$\phi\left(\cdot\,;\, \mu,\sigma^2\right)$ denotes a Gaussian density
with mean $\mu$ and variance $\sigma^2$. We place a normal-inverse
gamma prior on the parameter vector $\theta$, and we write
$(\mu,\sigma^2) \sim \mathrm{NIG}(m_0,\tau_0,\lambda_0,\gamma_0)$. In
other words, \begin{equation*}
	\mu\mid\sigma^2\sim \mathrm{N}\left(m_0,\frac{\sigma^2}{\tau_0}\right), \quad 1/\sigma^2 \sim \mathrm{Gamma}(\lambda_0,\gamma_0).
\end{equation*} Note that $\lambda_0$ and $\gamma_0$ are the shape
and rate parameters of the gamma distribution, respectively. Mixtures of
Gaussian kernels are one of the most used models for handling continuous
data and for approximating unknown densities; furthermore, they enjoy
good theoretical properties even when the kernel is misspecified
\citep{hjort2010}.

\begin{table}[t]
	\begin{tabularx}{\textwidth}{lll}
		Model   & Prior on the mixture weights                                     & Hyperprior  \\
		\toprule
		CAM \hspace{.5cm}    &     $\boldsymbol{\pi}\sim \mathrm{GEM}(\alpha)$                      &   $\alpha \sim\mathrm{Gamma}(h^{\alpha}_1,h^{\alpha}_2)$  \\
		&     $\boldsymbol{\omega}_k \sim \mathrm{GEM}(\beta), \:k \geq 1$      &   $\beta \sim\mathrm{Gamma}(h^{\beta}_1,h^{\beta}_2)$  \\\midrule
		FISAN   &     $\boldsymbol{\pi}\sim \mathrm{GEM}(\alpha)$                      &   $\alpha \sim\mathrm{Gamma}(h^{\alpha}_1,h^{\alpha}_2)$  \\
		&     $\boldsymbol{\omega}_k \sim \mathrm{Dirichlet}_L(b), \:  k =1,\ldots,K\qquad$   &       \\\midrule
		FSAN    &     $\boldsymbol{\pi}\sim \mathrm{Dirichlet}_K(a)$                           &      \\
		&     $\boldsymbol{\omega}_k \sim \mathrm{Dirichlet}_L(b), \:  k =1,\ldots, K$  &           \\
		\bottomrule 
	\end{tabularx}
	\caption{Summary of the implemented models along with their corresponding prior and hyperprior specifications over the mixture weights.}
	\label{tab::functions}
\end{table}

\subsection{Computational methods for posterior inference}
\label{sec::posterior_inference}

The models we have just introduced rely on two nested levels of
mixtures. Developing general and computationally efficient posterior
inference techniques for Bayesian mixture models is an ongoing research
topic. A particular challenge of these types of models is related to the
definition of the appropriate number of mixture components. Before
detailing the different approaches, we emphasize the distinction between
the concepts of \textit{clusters} and \textit{components}, which is
central to many modeling and computational strategies. Following
\citet{fs2011}, we denote as \emph{components} all the terms involved in
the mixture. In contrast, with \textit{clusters} we refer only to the
non-empty ones. Consequently, the number of mixture components ($K$
and $L$, in our formulation) may differ from the number of clusters in
a given sample.

DP mixtures and, in general, nonparametric mixtures assume that the
number of components is infinite, an aspect that requires particular
care when developing computational strategies. MCMC methods, especially
Gibbs sampling, are currently established as the standard methodology.
These algorithms are often distinguished into \textit{conditional} and
\textit{marginal} methods, depending on how they deal with the
nonparametric process. Here, we only focus on conditional methods and,
specifically, on the slice sampler of \citet{slice2}, as it provides a
good balance between computational efficiency and precision of the
approximation of the posterior to the true nonparametric process.
However, the growing demand for computational methods suitable for big
data analysis has recently prompted the development of VI algorithms for
nonparametric mixtures \citep{Blei2006}. VI tackles posterior estimation
as an optimization rather than as a simulation problem. Specifically, it
is based on minimizing the Kullback-Leibler divergence between the
elements of a class of simpler, manageable distributions and the true
posterior. The most straightforward framework is mean-field variational
inference, where some (or all) of the dependencies between the
parameters are ignored. The coordinate ascent variational inference
(CAVI) algorithm represents one of the most common methods, updating
each coordinate sequentially. This algorithm leads to the minimization
of the Kullback-Leibler divergence, or, equivalently, to the
maximization of the so-called evidence lower bound (ELBO)
\citep{Bishop2006, Blei2017}.

Finite mixtures, as the name suggests, assume instead that the number of
components is a finite integer. Recently, there has been a renewed
interest in models based on finite mixtures that do not require
knowledge of the number of components or clusters.
\citet{millerharrison2018} and \citet{fs2011}, for example, developed
frameworks for finite mixtures with a random number of components. Their
formulations allowed the derivation of straightforward strategies for
posterior inference despite the difficulties stemming from having a
varying-dimensional parameter space. A different interesting perspective
was investigated by \citet{Rousseau2011} and \citet{MalsinerWalli2016},
which showed how finite mixtures (with a fixed and known number of
components) can be used to estimate the number of clusters in a sample
by defining an adequate prior distribution on the mixing probabilities.
Specifically, \citet{Rousseau2011} showed how a mixture specification
based on a large, ``overfitting'' number of components (i.e., much
larger than the number of clusters expected in the data) and a Dirichlet
prior of parameter $a < \zeta/2$, with $\zeta$ the dimension of the
component-specific parameter, asymptotically concentrates on the true
mixture, \textit{a posteriori}. In other words, the posterior will empty
the extra components and allow for the estimation of the dimension of
the partition of the observed sample. An important strength of this
approach is that all available algorithms for finite mixture models can
be adapted to estimate these sparse mixtures. This approach, however,
can lead to shortcomings in practical applications. Fixing the dimension
of the Dirichlet distribution to very large values leads to an increased
computational cost in terms of memory usage and computing time. This
aspect can become prominent when using MCMC methods, especially if large
datasets are involved. Nevertheless, the simplicity of the underlying
theoretical framework allows for exploiting standard VI algorithms,
which dramatically reduces the related computational cost.

The proposed \pkg{sanba} package provides functions to straightforwardly
estimate the CAM, FISAN, and FSAN models using both MCMC and VI
approaches. The pseudocode of the implemented algorithms is provided in
the original papers. In particular, all MCMC approaches for DP mixtures
use the nested slice sampler of \citet{Denti2023}. VI methods for DP
mixtures are based on the nested extension of the CAVI algorithm
\citep{Blei2006}, detailed in \citet{SAN}. A standard Gibbs sampler and
CAVI algorithm with a fixed number of components are implemented for
finite mixtures, with default parameters set to induce sparsity and
infer the number of clusters.

\section{How to use sanba: an illustration}
\label{sec::package}

This section presents a detailed, step-by-step analysis of a multi-group
dataset using \pkg{sanba}. Recall that the package offers two estimation
procedures: MCMC and VI techniques. The former is better suited for
small to moderately sized datasets, providing more accurate estimates,
whereas VI is an approximate but faster method, ideal for scalable
inference on larger datasets.

Here, we illustrate how to use the main functions in \pkg{sanba} with
the aid of a simple simulated dataset, where the data are generated
according to a finite mixture of Gaussians. In particular, we consider
15,000 observations organized into 15 groups. The observations in each
group are generated according to one of three group-specific
distributions (the distributional clusters), $f_1(y)$, $f_2(y)$, and
$f_3(y)$, each sampled with probability $1/3$. According to our
model, each density is a mixture of normal distributions, and, in
particular, we consider the following bimodal mixtures with shared
atoms: \begin{equation}
	\begin{gathered}
		f_1(y) = \frac12 \phi\left(y\,;\, -5,1\right)+ \frac12 \phi\left(y\,;\, 0,1\right),\quad\quad
		f_2(y) = \frac12 \phi\left(y\,;\, -5,1\right)+ \frac12 \phi\left(y\,;\, 5,1\right),\\
		f_3(y) = \frac12 \phi\left(y\,;\, 0,1\right)+ \frac12 \phi\left(y\,;\, 5,1\right).
	\end{gathered}
\end{equation} Finally, we generate the observations by drawing 1000
independent samples from each group-specific distribution. We store the
observations in a vector \texttt{Y}, while the group labels are stored in
the vector \texttt{group}. The group assignment is encoded using integers
ranging from 1 to 15.

	\begin{verbatim}
		R> set.seed(100000)
		R> group <- sort(rep(1:15, 1000))
		R> distributional_cluster <- sample(1:3, size = 15, 
		+                                 replace = TRUE, prob = c(1/3, 1/3, 1/3))
		R> generate_obs <- function(size = 1000, dc){
			+   if(dc == 1) { atoms = c(-5, 0) }
			+   if(dc == 2) { atoms = c(-5, 5) }
			+   if(dc == 3) { atoms = c( 0, 5) }
			+   observational_cluster <- sample(atoms, size, 
			+                                  replace = TRUE, prob = c(0.5, 0.5))
			+   return(observational_cluster)
			+ }
		R> set.seed(250892)
		R> observational_cluster <- c(sapply(distributional_cluster, 
		+                                  function(x) generate_obs(dc = x) ))
		R> Y <- rnorm(length(observational_cluster), observational_cluster, 1)
		R> observational_cluster <- as.numeric( as.factor(observational_cluster) )
	\end{verbatim}

\begin{figure}[t!]
	\centering
	\includegraphics[width = \linewidth]{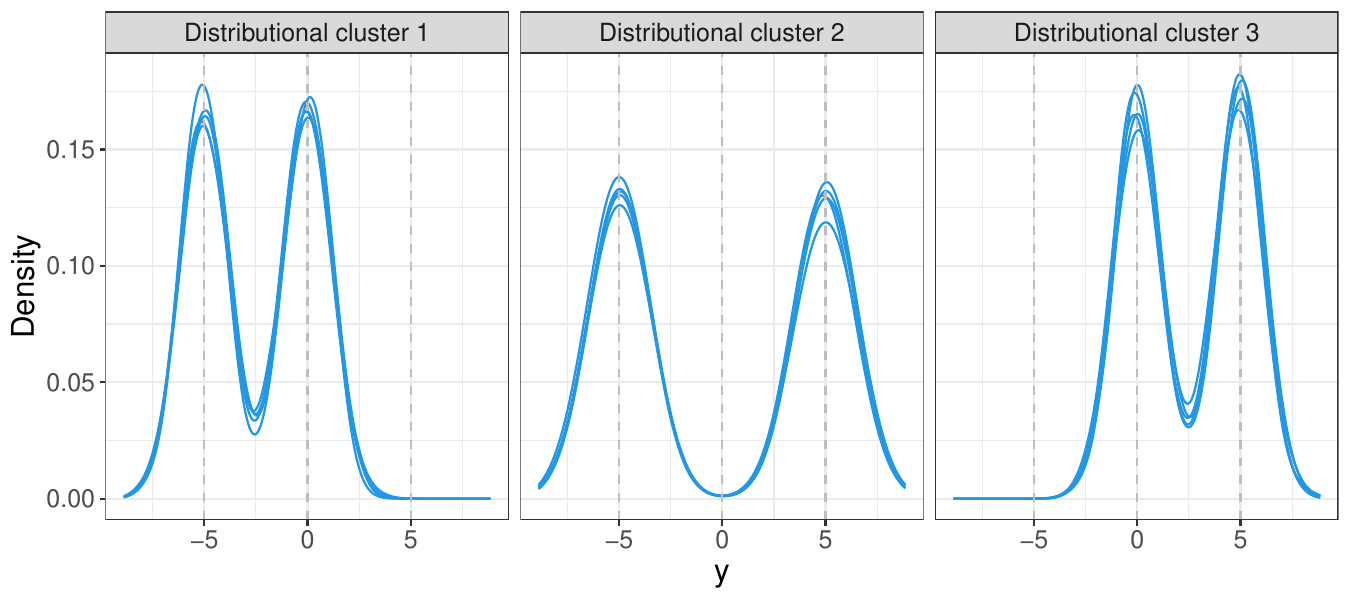}
	\caption{Kernel density estimates of the simulated data (one line for each group), stratified by their true distributional clusters.}
	\label{fig::descriptive_simu}
\end{figure}

The vectors \texttt{Y} and \texttt{group} are all we need to perform the
analyses. Figure\nobreakspace{}\ref{fig::descriptive_simu} visually
displays the empirical density estimates of each sample, divided by
distributional cluster.

Before proceeding, let us load the package:

	\begin{verbatim}
		R> library("sanba")
	\end{verbatim}

One can fit the CAM, FISAN, and FSAN models by invoking the functions
\texttt{fit\_CAM()}, \texttt{fit\_fiSAN()}, and \texttt{fit\_fSAN()},
respectively. These functions take three common arguments as input: the
data \texttt{y}, the group assignment \texttt{group}, and the estimation
method \texttt{est\_method}, which can be set to one of the elements of
\texttt{c("VI", "MCMC")}. Additionally, one can specify one
model-specific list of hyperparameters (\texttt{prior\_param}) and a
method-specific list of tuning parameters (\texttt{mcmc\_param} or
\texttt{vi\_param}), both of which are optional. These arguments are
summarized in Tables\nobreakspace{}\ref{tab::prior_param_summary}
and\nobreakspace{}\ref{tab::method_param_summary}. In the following
sections, we provide complete pipelines using both the MCMC and the VI
approaches, focusing on the FISAN model. The workflow for the other
nested models involves only minor adjustments in the specification of
the hyperparameters in the \texttt{prior\_param} list.

\begin{table}[t]
	\begin{tabularx}{\textwidth}{llX}
		Function          & \texttt{prior\_param}                   & Description \\
		\toprule
		\texttt{fit\_CAM()} &  \texttt{m0, tau0, lambda0, gamma0}     &  Parameters of the NIG distr. on $(\mu,\sigma^2)$ \\
		&  \texttt{hyp\_alpha1, hyp\_alpha2}      &  Parameters of the gamma distr. on $\alpha$ \\
		&  \texttt{hyp\_beta1, hyp\_beta2}        &  Parameters of the gamma distr. on $\beta$ \\
		&  \texttt{alpha}      &  Optional: to set a fixed $\alpha$ \\
		&  \texttt{beta}        & Optional: to set a fixed $\beta$ \\
		\midrule
		\texttt{fit\_fiSAN()} &  \texttt{m0, tau0, lambda0, gamma0}     &  Parameters of the NIG distr. on $(\mu,\sigma^2)$ \\
		&  \texttt{hyp\_alpha1, hyp\_alpha2}      &  Parameters of the gamma distr. on $\alpha$ \\
		&  \texttt{b\_dirichlet}        & Parameter $b$ of the observational Dirichlet \\
		&  \texttt{alpha}      &  Optional: to set a fixed $\alpha$ \\
		\midrule
		\texttt{fit\_fiSAN()} &  \texttt{m0, tau0, lambda0, gamma0}     &  Parameters of the NIG prior on $(\mu,\sigma^2)$ \\
		&  \texttt{a\_dirichlet}      &  Parameter $a$ of the distributional Dirichlet \\
		&  \texttt{b\_dirichlet}        & Parameter $b$ of the observational Dirichlet \\
		\bottomrule
	\end{tabularx}
	\caption{Summary of the model-specific prior parameters that can be passed to the functions \texttt{fit\_CAM()}, \texttt{fit\_fiSAN()}, and \texttt{fit\_fSAN()} via the \texttt{prior\_param} list.}
	\label{tab::prior_param_summary}
\end{table}

\begin{table}[t]
	\begin{tabularx}{\textwidth}{llX}
		\texttt{method}          &      \texttt{mcmc\_param}  or \texttt{vi\_param}         & Description \\
		\toprule
		\texttt{MCMC} &  \texttt{nrep, burn}     &  Integers, the number of total MCMC iterations, and the number of discarded iterations, respectively \\
		&  \texttt{maxL, maxK}     &  Upper bounds for the number of observational and distributional clusters, respectively \\
		&  \texttt{seed}        &  Integer to set the random seed for reproducibility\\
		&  \texttt{warmstart}        & Logical, if \texttt{TRUE}, the observationalcluster means are initialized using the k-means algorithm \\
		&  \texttt{verbose}      &  Logical, if \texttt{TRUE} progress messages are printed during sampling \\
		\midrule
		\texttt{VI} &  \texttt{maxL, maxK}     &  Upper bounds for the number of observational and distributional clusters, respectively \\
		&  \texttt{epsilon}      &  Tolerance that drives the convergence criterion adopted as the stopping rule \\
		&  \texttt{seed}        &  Integer to set the random seed for reproducibility\\
		&  \texttt{maxSIM}      &  Maximum number of iterations for the VI algorithm \\
		&  \texttt{warmstart}        & Logical, if \texttt{TRUE}, the observational cluster means are initialized using the k-means algorithm \\
		&  \texttt{verbose}      &  Logical, if \texttt{TRUE} progress messages are printed during optimization \\
		\bottomrule
	\end{tabularx}
	\caption{Summary of the method-specific parameters that can be passed to the functions \texttt{fit\_CAM()}, \texttt{fit\_fiSAN()}, and \texttt{fit\_fSAN()} via the \texttt{mcmc\_param} or \texttt{vi\_param} lists.}
	\label{tab::method_param_summary}
\end{table}

\subsection{MCMC inference with {sanba}}

The core function for performing posterior inference for the FISAN model via MCMC is \texttt{fit\_fiSAN()}, which should be called with the argument
\texttt{est\_method = "MCMC"}. The MCMC-specific settings are provided
through the \texttt{mcmc\_param} list (see
Table\nobreakspace{}\ref{tab::method_param_summary}), which allows the
user to control, among others, the number of iterations and the burn-in
period, the initialization strategy, and the random \texttt{seed} for
reproducibility. In the example below, we generate posterior samples
using 10,000 MCMC iterations, discarding the first half as burn-in. This
results in a retained chain of 5,000 samples. The FISAN model is
estimated using default prior parameters (i.e.,
$\alpha \sim \mathrm{Gamma}(1, 1)$ and $b = 1/\texttt{maxL}$). The
algorithm is initialized with a partition corresponding to five
observational clusters, specified via \texttt{nclus\_start = 5}:

	\begin{verbatim}
		R> res_mcmc <- fit_fiSAN(y = Y, group = group, est_method = "MCMC",
		+                       mcmc_param = list(nrep = 10000, burn = 5000,
		+                                         seed = 123, nclus_start = 5, 
		+                                         maxL = 50, maxK = 50) )
	\end{verbatim}

The returned output, \texttt{res\_mcmc}, is an object of class
\texttt{SANmcmc}. It contains the following components:

\begin{itemize}
	\item \texttt{model}: a string declaring the type of fitted model;
	\item \texttt{param}: a list containing the data vector \texttt{Y}, the corresponding group labels, and all the user-specified or default parameters;
	\item \texttt{sim}: a list containing the Markov chains of all sampled parameters;
	\item \texttt{time}: the computation time required to generate the \texttt{nrep} replications.
\end{itemize}

When printed, the object returns a concise overview of the fitting
procedure, including the total number of iterations, the burn-in period,
and the associated computational cost. A more detailed description can
be obtained via the \texttt{summary()} method, which reports the model and
parameter specifications, fitting outcomes, and summary statistics on
the inferred partitions:

	\begin{verbatim}
		R> summary(res_mcmc)
	\end{verbatim}
\begin{verbatim}
		
		MCMC results for fiSAN 
		-----------------------------------------------
		Model estimated on 15000 total observations and 15 groups 
		maxL: 50 - maxK: 50 
		Prior parameters (m0, tau0, lambda0, gamma0): ( 0, 0.01, 3, 2 ) 
		
		Size of the MCMC sample (after burn-in): 5000 
		Total MCMC iterations performed: 10000 
		Elapsed time: 3.723 mins 
		
		Number of observational and distributional clusters:
		OC DC
		Mean     9.162200  3
		Median   9.000000  3
		Variance 4.872466  0
	\end{verbatim}

To assess convergence, users can rely on the \texttt{plot()} function,
which displays traceplots and posterior density estimates of the
selected parameters. For example,
Figure\nobreakspace{}\ref{fig::traceplot} shows the resulting plots of
the mean parameters $\mu_l$, for $l=1,2,3$ in the top panels, and of
the DP concentration parameter $\alpha$ in the bottom panels (only
available if $\alpha$ is random). These plots can be obtained using
the following code:

	\begin{verbatim}
		R> plot(res_mcmc, param = "mu", trunc_plot = 3)
		R> plot(res_mcmc, param = "alpha")
	\end{verbatim}

\begin{figure}[t]
	\centering
	\includegraphics[width = \linewidth]{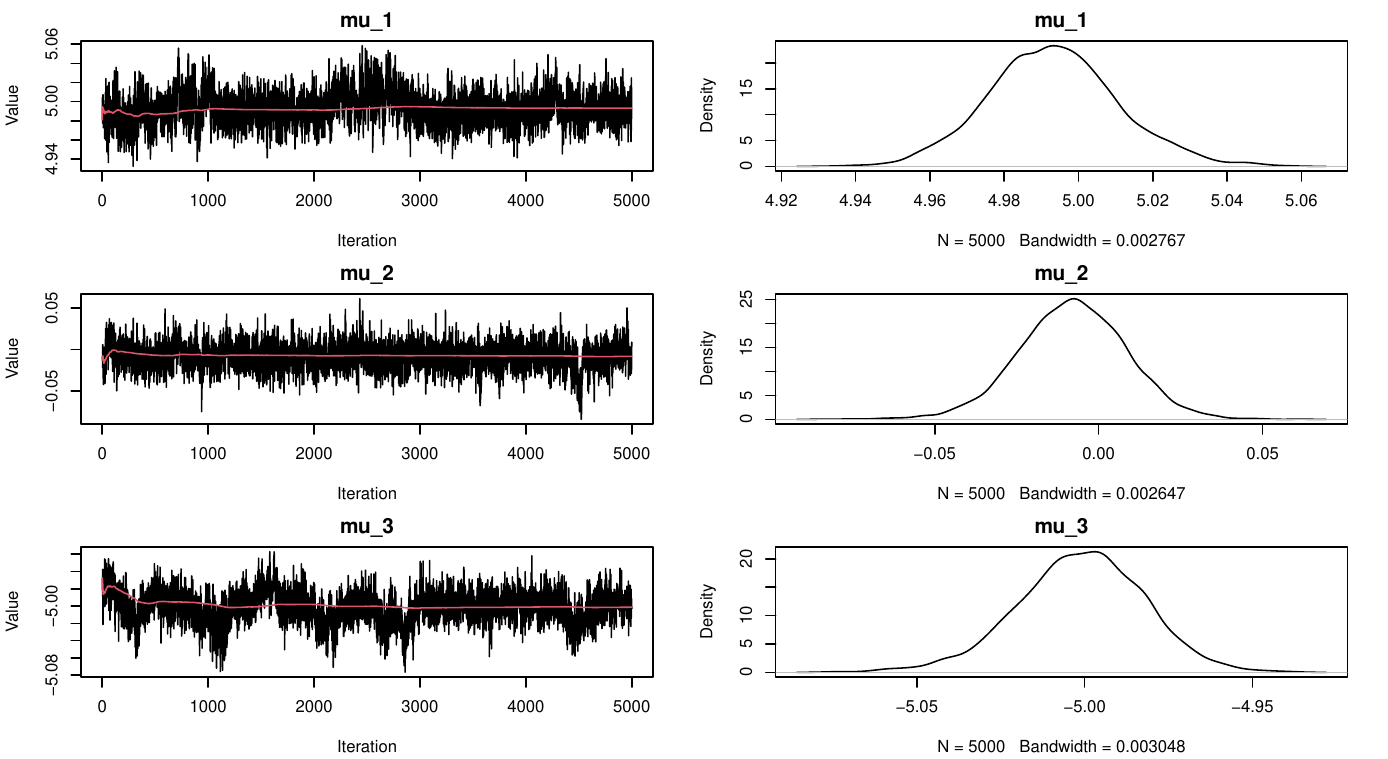}
	\includegraphics[width = \linewidth]{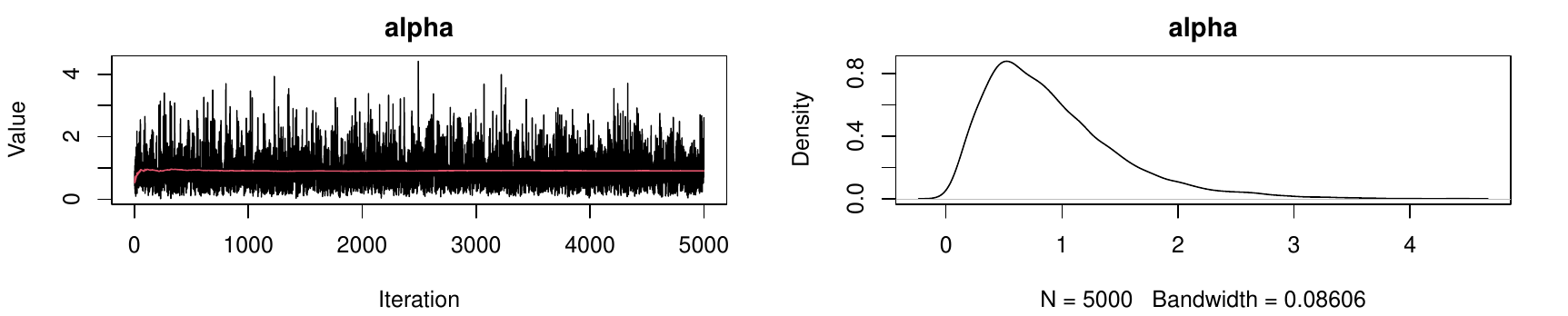}
	\caption{Result of the \texttt{plot()} method for the \texttt{SANmcmc} object \texttt{res\_mcmc} applied to the parameters \texttt{"mu"} (top three panels) and \texttt{"alpha"} (bottom panels). Left column: The traceplots (black lines) and the ergodic means (red lines). Right column: Posterior kernel density estimates.}
	\label{fig::traceplot}
\end{figure}

Other parameters that can be visualized through this function include
the cluster-specific variances (obtained by setting
\texttt{param = "sigma2"}), the number of clusters (\texttt{"num\_clust"}),
the distributional probabilities (\texttt{"pi"}), and, for the CAM model,
the observational concentration parameter $\beta$ (\texttt{"beta"}).

After assessing convergence, one can proceed with posterior inference.
We can compute a posterior estimate of the observational and
distributional partitions with a call to the function
\texttt{estimate\_partition()}, applied to the object of class
\texttt{SANmcmc}. The function employs the \texttt{salso()} algorithm of the
homonymous package\nobreakspace{}\citep{salso} to efficiently compute
the posterior point estimates. This method returns an object of class
\texttt{partition\_mcmc}, which is a list containing two elements:

\begin{itemize}
	\item \texttt{dis\_level}: a vector containing the DC assignment for each group;
	\item \texttt{obs\_level}: a dataset containing, in each row, the data point, its corresponding group, and the estimated DC and OC assignments.
\end{itemize}

The \texttt{summary()} method applied to this object displays the
estimated number of observational and distributional clusters. Although
\texttt{salso()} implements an efficient algorithm to find the optimal
partition, such a procedure is still computationally intensive,
especially with large datasets. To mitigate this, one can use the
optional argument \texttt{add\_burnin} to focus on even shorter MCMC
chains.

	\begin{verbatim}
		R> clusters <- estimate_partition(res_mcmc)
		R> summary(clusters)
	\end{verbatim}

	\begin{verbatim}
		Summary of the posterior observ. and distrib. partitions estimated via MCMC
		----------------------------------
		Number of estimated OCs: 3 
		Number of estimated DCs: 3 
		----------------------------------
	\end{verbatim}

\begin{figure}[t!]
	\centering
	\includegraphics[width = \linewidth]{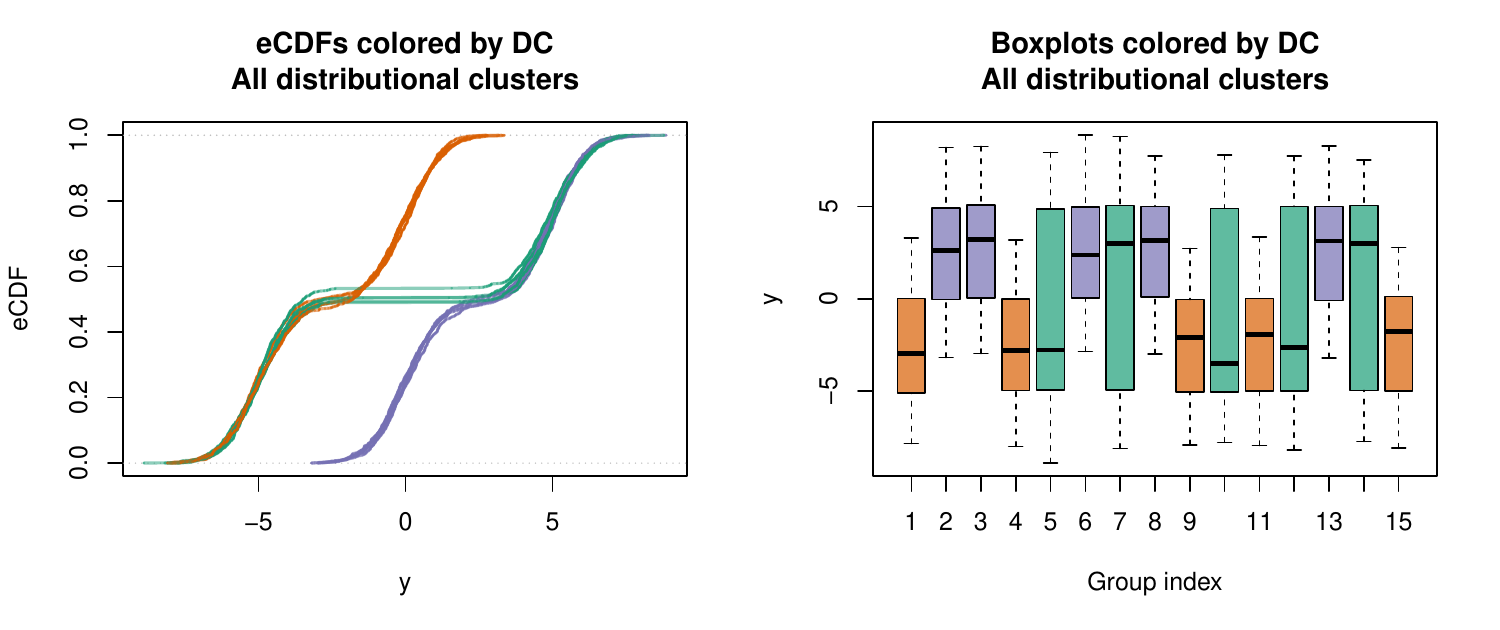}
	\includegraphics[width = \linewidth]{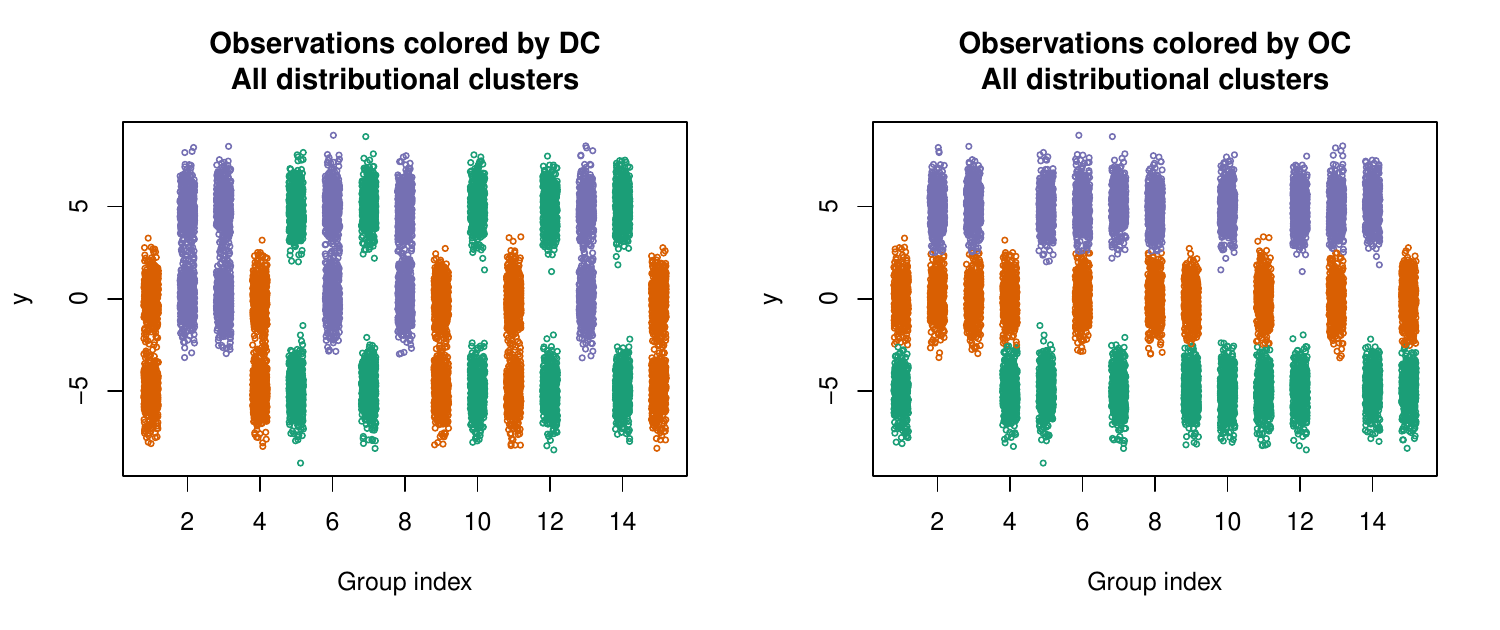}
	\caption{Result of the \texttt{plot()} method for the \texttt{partition\_mcmc} object \texttt{clusters}. Top-left: ECDFs of each group, colored by DC. Top-right: Boxplots of each group, colored by DC. Bottom: scatter plots with observations colored by DC and OC.}
	\label{fig::plot_mcmc}
\end{figure}

Finally, we can visually analyze the estimated partitions with a call to
the \texttt{plot()} function applied to a \texttt{partition\_mcmc} object
(Figure\nobreakspace{}\ref{fig::plot_mcmc}). One can choose between the
empirical cumulative distribution functions (ECDFs), using
\texttt{type = "ecdf"}, the boxplots of each group, colored by DC
(\texttt{type = "boxplot"}), and two scatter plots with observations
colored by DC and OC (\texttt{type = "scatter"}). The \texttt{alt\_palette}
option allows using an alternative palette to {\proglang{R}}{} base
colors; moreover, additional graphical parameters can be passed to
improve visualization.

	\begin{verbatim}
		R> plot(clusters, type = "ecdf", alt_palette = TRUE, lwd = 2)
		R> plot(clusters, type = "boxplot", alt_palette = TRUE)
		R> plot(clusters, type = "scatter", alt_palette = TRUE, cex = 0.5)
	\end{verbatim}

Thus, with just a few simple function calls, we have demonstrated a
complete analysis of synthetic nested data using an MCMC approach. We
voluntarily omitted a discussion of the technical details regarding the
use of a simulation approach (e.g., the length of the chains, the
assessment of convergence, and the label switching issue) since MCMC
methods are well-known and typically regarded as standard practice.

\subsection{Variational inference with sanba}

Posterior inference for the FISAN model via coordinate ascent
optimization is carried out using the \texttt{fit\_fiSAN()} function,
called with the argument \texttt{est\_method = "VI"}. The algorithm
iteratively maximizes the ELBO until convergence. In this context,
convergence is defined as the point at which the improvement in the ELBO
between two successive iterations falls below a user-specified threshold
\texttt{epsilon}.

Although this function can theoretically be used ``as is,'' it is
important to note that the CAVI algorithm is only guaranteed to find
local optima. As such, the final solution is sensitive to the choice of
initialization. For this reason, it is customary, and strongly
recommended, to perform multiple runs of the algorithm using different
starting values and to retain the run that achieves the highest ELBO
value to perform inference. The \texttt{fit\_fiSAN()} function, along with
the other model-fitting routines, facilitates this process via the
\texttt{n\_runs} argument, which executes multiple optimization attempts.
While the function stores the ELBO trajectories for all runs, it saves
only the optimized parameter values corresponding to the run that
attained the best ELBO. Algorithm-specific settings can be passed via
the \texttt{vi\_param} list (see
Table\nobreakspace{}\ref{tab::method_param_summary}), analogous to the
\texttt{mcmc\_param} list used in MCMC-based fitting. The function returns
an object of class \texttt{SANvi}, which includes all relevant
information for downstream analysis.

To illustrate its use, we fit the FISAN model to the synthetic dataset
introduced before. The code below performs 50 independent runs, with
convergence defined as an ELBO increment smaller than 0.01. The upper
bounds for the number of distributional and observational clusters are
set to 20 and 25, respectively. Regarding the model specification, we
assume a random concentration parameter $\alpha$ for the
distributional Dirichlet Process, assigning it a Gamma prior with both
shape and rate equal to 3. For the observational-level Dirichlet
distributions, the concentration parameter is fixed at 0.001.

\begin{verbatim}
		R> res_vi <- fit_fiSAN(y = Y, group = group, est_method = "VI",
		+                     prior_param = list(hyp_alpha1 = 3, hyp_alpha2 = 3,
		+                                        b_dirichlet = 0.001),
		+                     vi_param = list(seed = 2508,
		+                                     maxL = 25, maxK = 20,                
		+                                     epsilon = 0.01, n_runs = 50))
\end{verbatim}

The output of the CAVI optimization that best fits the data is contained
in the object \texttt{res\_vi}, of class \texttt{SANvi}. It includes the
following elements:

\begin{itemize}
	\item \texttt{model}: a string indicating the fitted model;
	\item \texttt{param}: a list containing the data vector \texttt{Y}, the corresponding group labels, and all supplied parameters;
	\item \texttt{sim}: a list containing the optimized variational parameters;
	\item \texttt{time}: the total computing time required to reach convergence across all \texttt{n\_runs}.
\end{itemize}

Similar to the MCMC case, additional tailored functions are available to
help the user explore the results of the fitted model. Specifically,
\texttt{print()}, \texttt{plot()}, and \texttt{summary()} methods are
available for objects of class \texttt{SANvi}. These functions return
similar inferences to their MCMC counterparts; however, the specific
quantities differ due to the different nature of the computational
approach, as detailed below.

A call to \texttt{summary()} produces a detailed overview of the
optimization: the fitted model type, the final ELBO value, the number of
iterations until convergence, the total elapsed time, and a summary of
the estimated cluster partitions.

	\begin{verbatim}
		R> summary(res_vi)
	\end{verbatim}
	\begin{verbatim}
		
		Variational inference results for fiSAN 
		-----------------------------------------------
		Model estimated on 15000 total observations and 15 groups 
		maxL: 25 - maxK: 20 
		Prior parameters (m0, tau0, lambda0, gamma0): ( 0, 0.01, 3, 2 ) 
		
		Threshold: 0.01 
		ELBO value: -17870.477 
		Best run out of 50 
		Convergence reached in 928 iterations
		Elapsed time: 18.755 secs 
		
		Number of observational and distributional clusters:
		OC DC
		Estimate  3  3
	\end{verbatim}

\begin{figure}[t!]
	\centering
	\includegraphics[width = \linewidth]{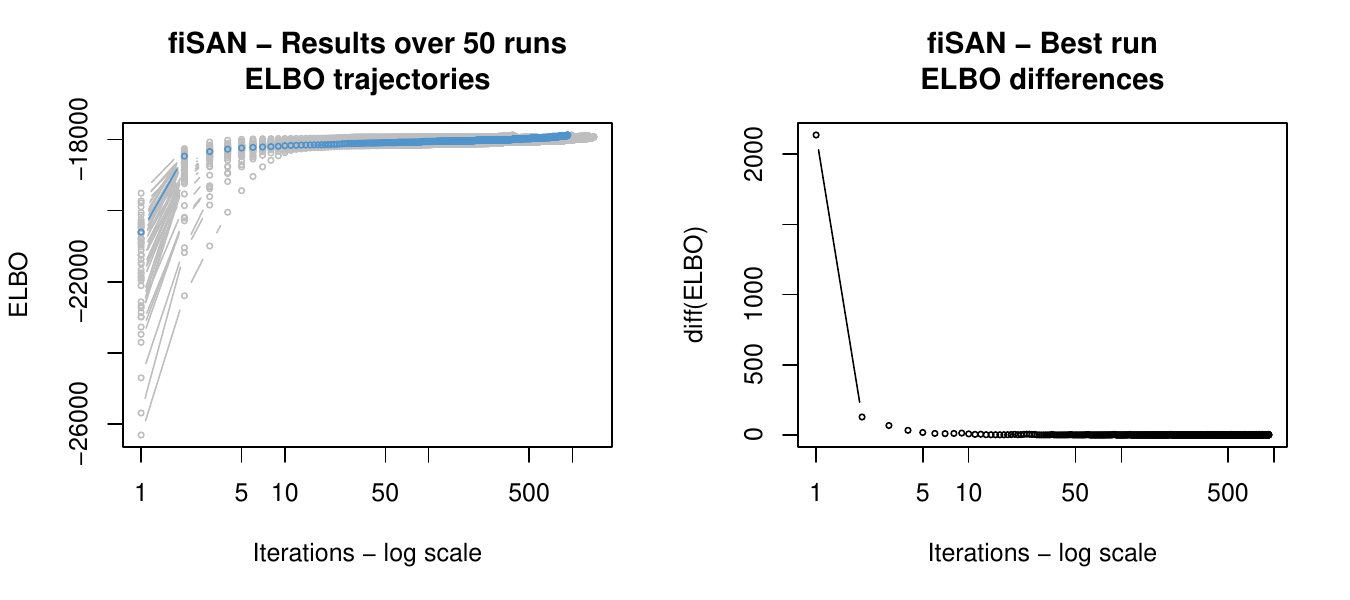}
	\caption{Result of the \texttt{plot()} method for the \texttt{SANvi} object \texttt{res\_vi}. Left panel: The grey lines show the 50 ELBO trajectories. The best run is highlighted in blue. Right panel: ELBO increments of the best run.}
	\label{fig::plot_simu_vi}
\end{figure}

The \texttt{plot()} method, when applied to a \texttt{SANvi} object with

	\begin{verbatim}
		R> plot(res_vi)
	\end{verbatim}

enables a visual assessment of the optimization process. Unlike in
MCMC-based inference, where convergence is assessed using trace plots,
VI relies on monitoring the ELBO evolution.
Figure\nobreakspace{}\ref{fig::plot_simu_vi} illustrates the result of a
call to the \texttt{plot} method applied to \texttt{res\_vi}. The left panel
displays the ELBO trajectories for all optimization runs, with the best
run highlighted in blue. The right panel shows the corresponding ELBO
increments, defined as $\Delta_{ELBO}(h) = ELBO_{h} - ELBO_{h-1}$ for
all iterations $h > 1$.

Turning to posterior inference, we estimate the observational and
distributional cluster assignments using the \texttt{estimate\_partition()}
function. The output is an object of class \texttt{partition\_vi}, which
shares the same structure as the \texttt{partition\_mcmc} object introduced
earlier. This object has dedicated \texttt{print()}, \texttt{summary()}, and
\texttt{plot()} methods. The usage and output of these functions are
analogous to those detailed for \texttt{SANmcmc} objects. Hence, we do not
discuss it again here.

Since the VI procedure provides a point estimate and does not suffer
from the label-switching issue affecting MCMC methods, we can safely
perform inference on the latent random measures $G_k^*$'s in a
relatively simple manner. This can be achieved using the function
\texttt{estimate\_G()}, which estimates the posterior means, variances, and
mixture weights characterizing the OCs within each DC. More
specifically, the function returns an object of class \texttt{SANvi\_G},
which is a well-structured data frame containing all relevant
information. When printed or summarized, it shows only the OCs with a
posterior mixture weight larger than 0.01, thus avoiding overly verbose
outputs caused by negligible clusters. This threshold can be adjusted
via the \texttt{thr} argument.

	\begin{verbatim}
		R> Gs <- estimate_G(res_vi)
		R> summary(Gs)
	\end{verbatim}
	\begin{verbatim}
		Estimated random measures via VI
		----------------------------------
		Atoms with posterior weight > 0.01 
		Number of detected DCs: 3 
		----------------------------------
		
		Distributional cluster # 1 
		post_mean post_var post_weight
		2    -4.998    1.004       0.494
		1    -0.008    1.028       0.506
		
		Distributional cluster # 2 
		post_mean post_var post_weight
		2    -4.998    1.004       0.505
		3     4.990    0.991       0.495
		
		Distributional cluster # 3 
		post_mean post_var post_weight
		1    -0.008    1.028       0.491
		3     4.990    0.991       0.509
	\end{verbatim}

This object can also be visualized using \texttt{plot()}, which returns
the graph shown in Figure\nobreakspace{}\ref{fig::plot_aw1}. This plot
displays the posterior densities of the estimated distributional atoms,
where vertical bars indicate the posterior means and bar heights
represent the associated mixture weights. The user may set the
\texttt{DC\_num} argument to display specific DCs, a useful feature for
large datasets with many clusters. For instance, the following code
plots only DCs 1 and 3:

	\begin{verbatim}
		R> plot(Gs, DC_num = c(1,3))
	\end{verbatim}

\begin{figure}[t!]
	\centering
	\includegraphics[width = \linewidth]{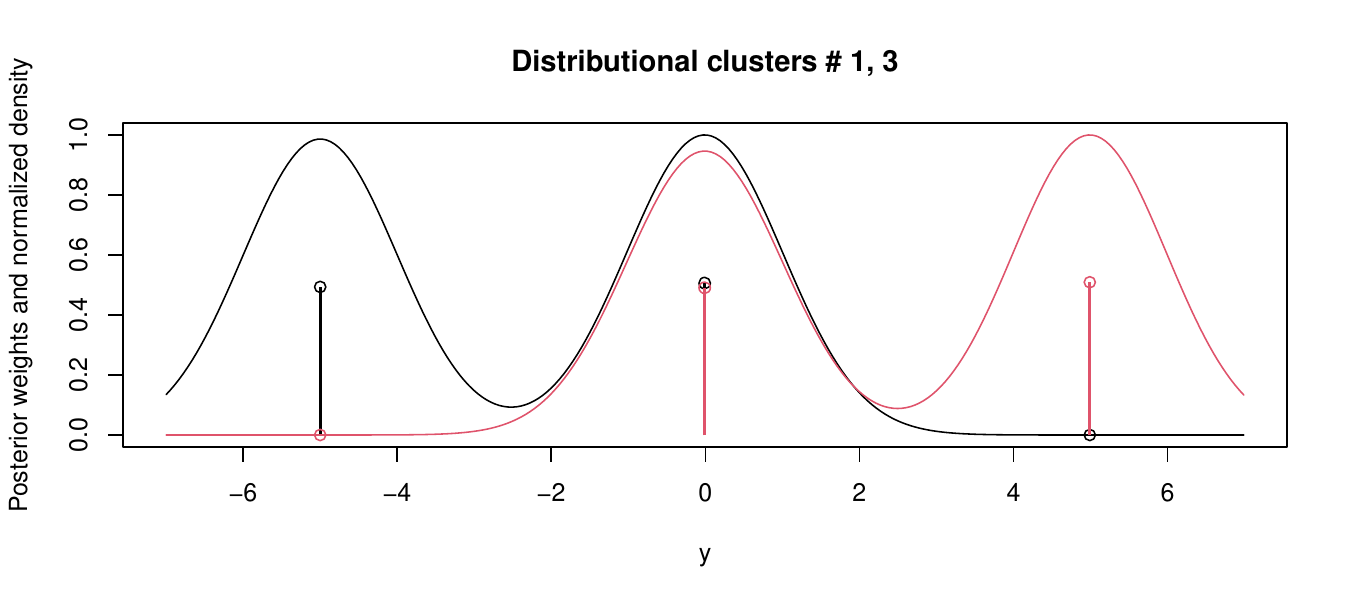}
	\caption{Posterior density estimates for the selected DCs (here, 1 and 3). The positions of the vertical bars indicate the posterior means, and their heights represent the assigned mixture weights.}
	\label{fig::plot_aw1}
\end{figure}

In this subsection, we have detailed a complete analysis workflow based
on the VI algorithm for the FISAN model as implemented in the
\pkg{sanba} package.\\

Finally, we emphasize that not all utility functions included in
\pkg{sanba} have been discussed here. For instance, several accessor
functions are available to retrieve specific components of the model
output, including \texttt{get\_model()}, \texttt{get\_param()},
\texttt{get\_sim()}, \texttt{get\_time()}, and, specifically for variational
inference, \texttt{get\_seed\_best\_run()}. Additional tools are also
provided to explore the estimated partitions, such as the computation of
posterior similarity matrices via \texttt{compute\_psm()} (for MCMC
output), the extraction of cluster number distributions via
\texttt{number\_clusters()}, and the visualization of variational cluster
assignment probabilities via \texttt{plot\_vi\_allocation\_prob()}.

In the next section, we present a real-data analysis in which we cluster
beers according to the distributions of their review scores, using a
popular high-dimensional dataset freely available online.

\section{Application to the beer ratings dataset}
\label{sec:beer}

The major advantage of a VI approach over standard MCMC methods is its
computational efficiency and scalability to large sample sizes. To
demonstrate how the \pkg{sanba} package allows a straightforward
application of nonparametric nested priors to real-world, large-scale
data, we consider the \texttt{Beer Reviews} data collected from the
website \texttt{Beer Advocate} and openly available on
Kaggle\footnote{https://www.kaggle.com/datasets/rdoume/beerreviews}. The
raw dataset comprises approximately 1.5 million reviews of hundreds of
beers, collected from a large number of users. Each review comprises the
ratings (on a discrete scale from 0 to 5) for five key beer
characteristics: \emph{appearance}, \emph{aroma}, \emph{palate} (i.e.,
mouthfeel), \emph{taste}, and \emph{overall} quality. In addition, the
dataset includes two descriptive variables: beer type and alcohol
content (\%ABV). This dataset addresses the question of which beers are
most appreciated by consumers. For each beer, a large number of reviews
are available: summarizing them with a single number (e.g., the mean or
the median rating) would not allow capturing the potential heterogeneity
and variability among individual preferences. A more comprehensive
analysis takes into account the whole distribution of the users' ratings
for each beer.

\begin{figure}[t]
	\centering
	\includegraphics[width = \linewidth]{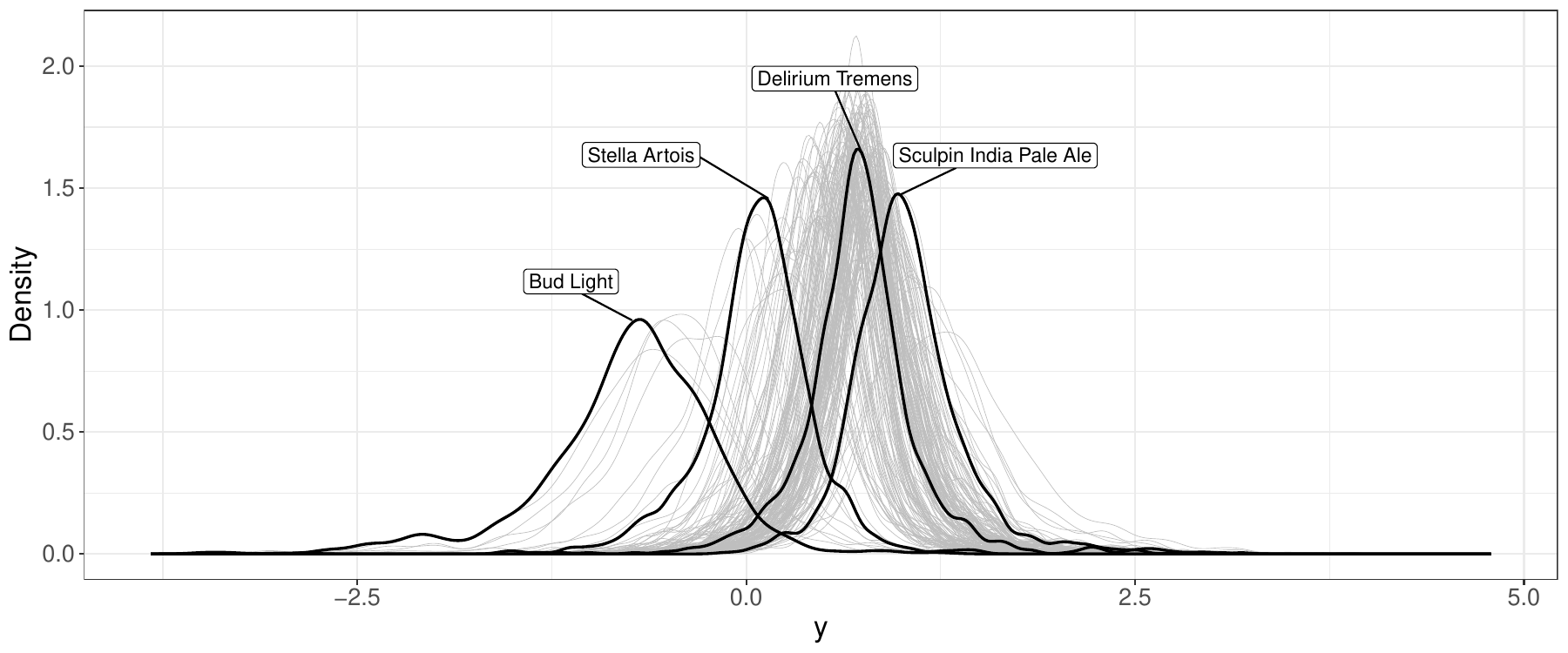}
	\caption{Kernel density estimates of the scores for the 170 beers contained in the dataset. Four different beers are highlighted to showcase the distributional differences that may occur across groups.}
	\label{fig::descriptive}
\end{figure}

In this application, we decided to focus on the general appreciation of
each beer and not on a particular aspect. For this reason, we computed a
continuous summary index of the five ratings. Specifically, we first
computed the average of the five indicators for each individual review.
Then, these scores were normalized and mapped onto the real line using
the inverse distribution function of a normal random variable. In the
following, this summary index is denoted as \texttt{score}. The code for
these preliminary preprocessing steps is available as Supplementary
Material. Moreover, we focused only on the beers that received a number
of reviews between 1000 and 2000 to avoid over- or under-represented
beers. After this filtering, we are left with 227,550 ratings (the
observations) over 170 beers (the groups), a challenging dimension for
most MCMC methods. To visualize the data,
Figure\nobreakspace{}\ref{fig::descriptive} displays the kernel density
estimates of the score distribution stratified by beers. In the plot, we
highlighted four famous beers: \texttt{Bud Light}, \texttt{Stella Artois},
\texttt{Delirium Tremens}, and \texttt{Sculpin IPA} to provide an example of
the heterogeneity of the distribution of the scores across groups.

We import and store the preprocessed dataset into the
\texttt{allbeers\_rating} dataset. Then, we save the computed scores and
group indexes into the vectors \texttt{beer\_ratings} and
\texttt{beer\_groups}.

	\begin{verbatim}
		R> allbeers_rating <- readRDS("RDS_Section4/preproc_beer.RDS")
		R> beer_ratings <- allbeers_rating$qnorm_rating
		R> beer_groups <- allbeers_rating$beer_id
	\end{verbatim}

We are now ready to study the latent nested clustering structure that
can be estimated from our observations.

\subsection{Segmenting beers into rating tiers}

We aim to segment the beers based on similarities in their score
distribution. This step can be seen both as an exploratory tool to
reduce the complexity and dimension of the data and as the final goal of
the analysis. In the latter case, the characteristics of the beers in
each distributional cluster can be analyzed to understand what specific
features drive the users' appreciation.

In both cases, the first step is the estimation of the model: this is
done by simply calling the function \texttt{fit\_fiSAN()}, as already seen
in the previous section. Specifically, we run the algorithm from 50
different starting points to fit the FISAN model. The model fitting was
launched on a desktop
computer\footnote{Intel Core i7-14700K \texttt{Ubuntu 22.04.5 LTS} workstation with 32 GB of RAM}.

	\begin{verbatim}
		R> fisan_beers <- fit_fiSAN(y = beer_ratings, group = beer_groups,
		+                          prior_param = list(b_dirichlet = 0.001),
		+                          vi_param = list(seed = 2508, maxL = 30, maxK = 20,
		+                                          epsilon = .01, n_runs = 50) )
		R> summary(fisan_beers)
	\end{verbatim}

	\begin{verbatim}
		
		Variational inference results for fiSAN 
		-----------------------------------------------
		Model estimated on 227550 total observations and 170 groups 
		maxL: 30 - maxK: 20 
		Prior parameters (m0, tau0, lambda0, gamma0): ( 0, 0.01, 3, 2 ) 
		
		Threshold: 0.01 
		ELBO value: 141071.305 
		Best run out of 50 
		Convergence reached in 981 iterations
		Elapsed time: 6.186 mins 
		
		Number of observational and distributional clusters:
		OC DC
		Estimate 13 10
	\end{verbatim}

\begin{figure}[t]
	\centering
	\includegraphics[width = \linewidth]{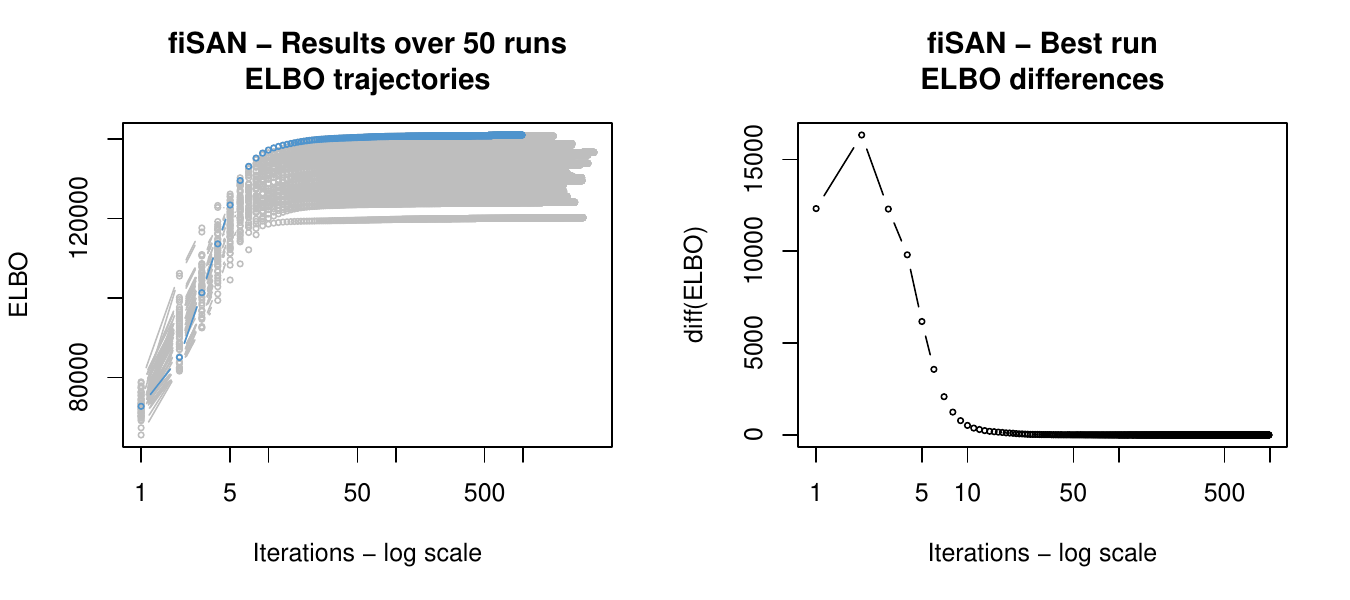}
	\caption{Left panel: The trajectories of the ELBO values for the 250 runs performed to cluster the beer ratings via VI. The best run is highlighted in blue. Right panel: ELBO increments of the best run.}
	\label{fig::elbobeers}
\end{figure}

The best run of the algorithm reached convergence in 981 iterations,
with a computing time of about six minutes. We can assess the 50 ELBO
trajectories with the \texttt{plot()} function to ensure that the fitting
phase was completed without issues. We report the results in
Figure\nobreakspace{}\ref{fig::elbobeers}.

	\begin{verbatim}
		R> plot(fisan_beers)
	\end{verbatim}

A call to the \texttt{estimate\_parition()} method reveals that the model
estimated 13 OCs and 10 DCs.

	\begin{verbatim}
		R> beers_clust <- estimate_partition(fisan_beers)
		R> summary(beers_clust)
	\end{verbatim}
	\begin{verbatim}
		Summary of the posterior observ. and distrib. partitions estimated via VI
		----------------------------------
		Number of estimated OCs: 13 
		Number of estimated DCs: 10 
		----------------------------------
	\end{verbatim}

We also extract the posterior estimates of the group-specific
distributions and save them in the following object:

	\begin{verbatim}
		R> beers_dists <- estimate_G(fisan_beers)
	\end{verbatim}

To better illustrate the richness and versatility of these outputs, here
we take advantage of the \pkg{ggplot2} package \citep{ggplotref} to
create more faceted plots (the code to produce them is deferred to the
Supplementary Material).

\begin{figure}[t]
	\centering
	\includegraphics[width = \linewidth]{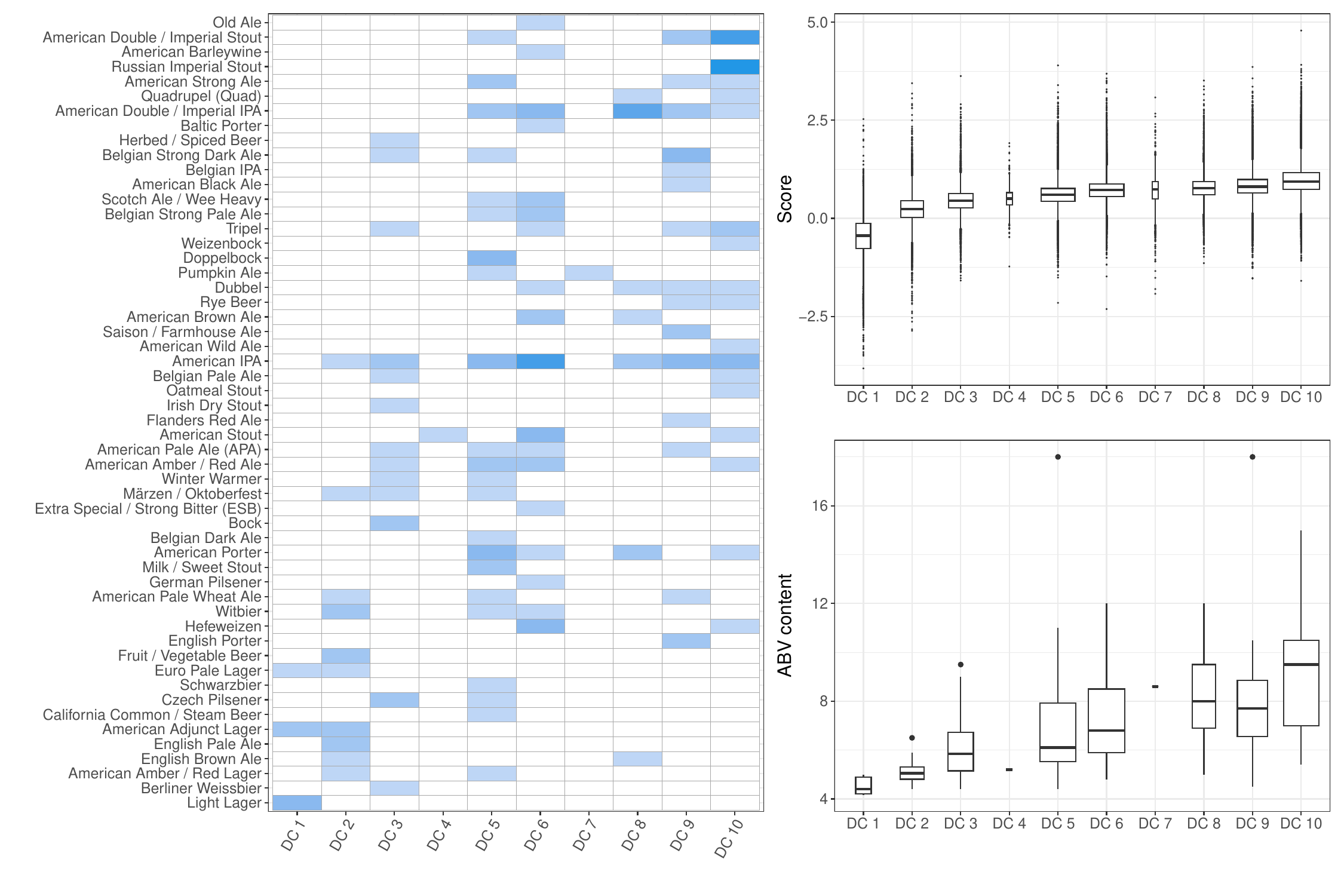}  
	\caption{Left panel: Heatmap of the contingency table showing the distribution of beer styles across DCs. Rows represent beer styles ordered by median ABV, and columns correspond to DCs. The color intensity reflects the number of beers of each style assigned to each DC.  
		Top-right panel: Distribution of scores within each DC, sorted in ascending order.  
		Bottom-right panel: Distribution of alcohol content (\%ABV) across DCs, sorted by increasing appreciation.}
	\label{fig::score_all}
\end{figure}

The content of \texttt{beers\_dists} can be used to analyze the
distributional clusters. To improve interpretability, we sorted the DCs
according to their sample median scores; thus, the DCs can be
interpreted as \emph{tiers of appreciation}, with DC 1 comprising the
least appreciated beers and DC 10 the users' favorites. The top-right
panel of Figure\nobreakspace{}\ref{fig::score_all} shows the
distribution of the scores of each DC: it is evident how the
appreciation increases with the DCs, confirming their interpretation.

\begin{figure}[t]
	\centering
	\includegraphics[width = \linewidth]{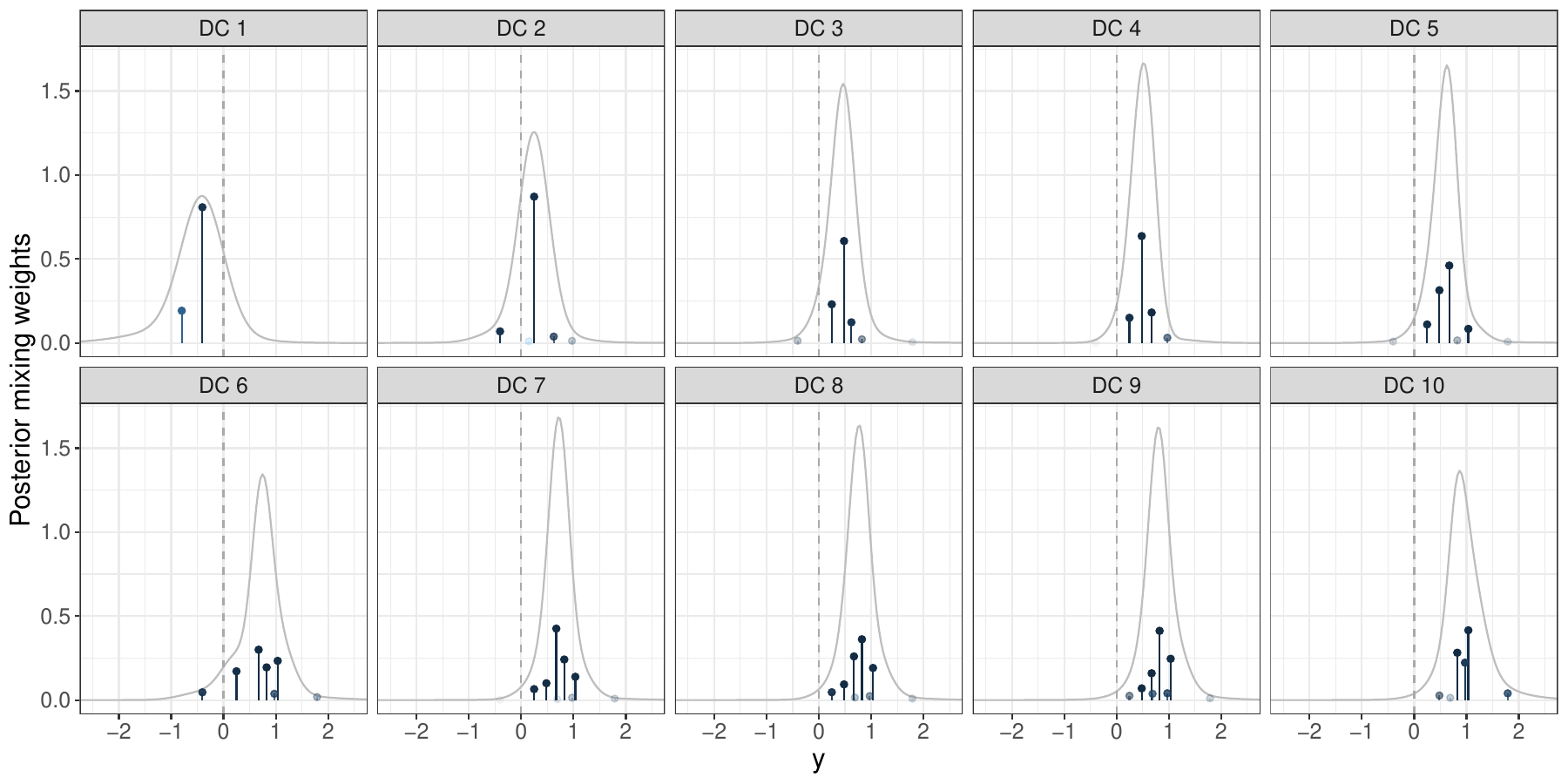}  
	\caption{Posterior density estimate of each DC (gray lines). Points correspond to the active observational atoms $\hat{\mu}_l$, their heights and color intensities are proportional to the estimated weights $\hat{\omega}_{l,k}$. }
	\label{fig::atomsbeers}
\end{figure}

However, \texttt{beers\_dists} allows obtaining more refined and
informative quantities: for example, we can recover and plot the
distributional atoms, i.e., the mixtures characterizing the score
distribution of the beers in the same distributional cluster. Recall
that the generic distributional atom $k$, for $k=1,\dots,K$, has the
following expression: \begin{equation}
	G^*_k(\cdot) = \sum_{l=1}^L \omega_{l,k} \, \delta_{\left(\mu_l,\sigma^2_l\right)}(\cdot).
	\label{eq:mix_recall}
\end{equation} Figure\nobreakspace{}\ref{fig::atomsbeers} shows, for
each DC, the estimated density
$\hat{f}_k(y) = \sum_{l=1}^L \hat{\omega}_{l,k}\phi(y\mid\hat{\mu}_l,\hat{\sigma}^2_l )$.
In addition to the estimated density, each panel shows the estimated
means $\hat{\mu}_l$ of the Gaussian kernels involved in the mixture.
Only the ``active'' atoms are displayed for each DC (i.e., the atoms
with non-negligible weight $\hat{\omega}_{l,k}$) to highlight the
distributional differences captured by each cluster. Notably, the
ordering of the DCs highlights how the active atoms gradually move from
the lowest to the highest values.

As already mentioned, the dataset also provides information about the
alcoholic content of each beer. It is then interesting to investigate
whether the \%ABV is related to the estimated partition. The
bottom-right Figure\nobreakspace{}\ref{fig::score_all} shows the
distribution of the beers' \%ABV across DCs, which are again sorted to
have increasing median scores. It is noticeable how the level of
appreciation has an evident positive association with the content of
alcohol. Indeed, except for DC 4 and DC 7, two singletons containing
only one beer, a higher score is always associated with a higher \%ABV.

To better understand this behavior, we can take a closer look at the
characteristics of four distinctive clusters: DC 1, which corresponds to
the cluster of the least appreciated beers; DC 8, which contains highly
appreciated beers (we do not study DC 10, corresponding to the top
beers, due to its large size to avoid burdensome outputs and graphics);
and DCs 4 and 7, which are the ``outliers'' in terms of the score
distributions. Table\nobreakspace{}\ref{tab::beers} reports the names,
styles, median score, and alcoholic content of the beers in these DCs.
We notice that all the beers in DC 1 are Lagers, which are generally
light, simple, and drinkable. DC 8 instead contains a variety of beer
styles, mainly IPAs or American Ales (but no Lager); what they all have
in common is a more peculiar and robust flavor. DCs 4 and 7 contain very
peculiarly flavored beers, made with pumpkin and chicory.

A plausible interpretation is to regard the alcohol content as a proxy
of the ``complexity'' of a beer, with this characteristic being the key
driver of user ratings. It is reasonable to assume that the users
assigning scores are avid beer enthusiasts. As such, they may prefer
robust and distinctive beers over more conventional, ``simpler''
options. This concept is strongly connected with the beer style; hence,
the homogeneity of the beer style within each DC.

\begin{table}[t]
	\footnotesize
	\begin{tabular}{llccc}
		DC & Beer name & Beer style & Median score & \%ABV \\
		\toprule
		&\texttt{Three Philosophers Belgian S. B.} & Quadrupel (Quad) & 1.129 & 9.80\\
		&\texttt{Double Simcoe IPA} & American Double / Imperial IPA & 1.091 & 9.00\\
		&\texttt{Black Butte Porter} & American Porter & 0.916 & 5.20\\
		&\texttt{Hop Stoopid} & American Double / Imperial IPA & 0.835 & 8.00\\
		&\texttt{Founders Double Trouble} & American Double / Imperial IPA & 0.813 & 9.40\\
		8&\texttt{Anchor Porter} & American Porter & 0.763 & 5.60\\
		&\texttt{Burton Baton} & American Double / Imperial IPA & 0.760 & 10.00\\
		&\texttt{Smuttynose IPA "Finest Kind"} & American IPA & 0.719 & 6.90\\
		&\texttt{Titan IPA} & American IPA & 0.719 & 7.10\\
		&\texttt{Westmalle Trappist Dubbel} & Dubbel & 0.682 & 7.00\\
		&\texttt{Samuel Smith's Nut Brown Ale} & English Brown Ale & 0.609 & 5.00\\
		&\texttt{Palo Santo Marron} & American Brown Ale & 0.405 & 12.00\\
		&\texttt{Unearthly} & American Double / Imperial IPA & 0.386 & 9.50\\
		\midrule
		7&\texttt{Pumking} & Pumpkin Ale & 1.311 & 8.60\\
		\midrule
		4&\texttt{Chicory Stout} & American Stout & 0.488 & 5.20\\
		\midrule
		&\texttt{Corona Extra} & American Adjunct Lager & 0.546 & 4.60\\
		&\texttt{Bud Light} & Light Lager & 0.215 & 4.20\\
		1&\texttt{Miller Lite} & Light Lager & -0.088 & 4.17\\
		&\texttt{Heineken Lager Beer} & Euro Pale Lager & -0.130 & 5.00\\
		&\texttt{Budweiser} & American Adjunct Lager & -0.219 & 5.00\\
		&\texttt{Coors Light} & Light Lager & -2.023 & 4.20\\
		\bottomrule
	\end{tabular}
	\label{tab::beers}
	\caption{Name, style, median score, and alcoholic content of the beers assigned to the top tier (DC 8, 13 beers), outlying distributions (DCs 4 and 7), and to the bottom tier (DC 1, 6 beers).}
\end{table}

To provide a broader perspective on the relationship between beer style
and DC assignment, in the left panel of
Figure\nobreakspace{}\ref{fig::score_all} we analyze the number of beers
with a specific beer style in each DC. The rows list beer styles,
ordered by median \%ABV, and the columns correspond to DCs. The
intensity of each cell reflects the number of beers of a given style in
each cluster. This visualization further confirms the previously
observed trend: lighter beer styles are predominantly found in the
lower-scoring DCs, whereas stronger, more complex styles appear in
higher-rated clusters.

\section{Summary and discussion}
\label{sec:conclusions}

In this paper, we have introduced the main features of \pkg{sanba}, an
\proglang{R} package that provides state-of-the-art MCMC and VI
algorithms for estimating the nested mixture models recently developed
by \citet{Denti2023} and \citet{SAN}. The package offers a compact suite
of high-level, user-friendly functions and methods, built upon efficient
and optimized low-level routines implemented in \proglang{C++}. We
remark that \pkg{sanba} also includes several additional functions not
covered in this paper; interested readers are referred to the package
documentation for further details.

To the best of our knowledge, \pkg{sanba} is currently the only
{\proglang{R}}{} package that enables both observational and
distributional clustering of nested datasets using Bayesian nested
mixture models with MCMC and VI estimation procedures. That said, there
remains considerable room for further development. A key future
extension involves incorporating alternative mixture kernels to enable
applications beyond the real line. For instance, mixtures of gamma
distributions could accommodate non-negative data, while Poisson or
rounded-Gaussian kernels \citep{tony} may be suited to modeling count
data. Another major enhancement would be the implementation of
multivariate kernels, allowing for the analysis of multivariate nested
structures. Incorporating more advanced VI strategies is also an
important direction. In particular, the current mean-field approach may
prove suboptimal when strong dependencies exist among variables.
Incorporating structured variational approximations or stochastic
variational inference methods would enhance scalability and enable the
application of these models to much larger datasets.

\bibliographystyle{jss}
\bibliography{refs}

\end{document}